\documentclass[epj]{svjour}

\usepackage{graphicx}
\usepackage{amsmath}

\begin{document}


\title{Synchronization of rotating helices by hydrodynamic interactions}
\author{Michael Reichert \and Holger Stark}
\institute{Fachbereich Physik, Universit\"at Konstanz,
  D-78457 Konstanz, Germany}
\mail{michael.reichert@uni-konstanz.de}
\date{Version: August 2, 2005}

\abstract{
%
%
Some types of bacteria use rotating helical flagella to swim.
The motion of such organisms takes place in the regime of low Reynolds
numbers where viscous effects dominate and where the dynamics is governed
by hydrodynamic interactions. Typically, rotating flagella form
bundles, which means that their rotation is synchronized.
The aim of this study is to investigate whether hydrodynamic
interactions can be at the origin of such a bundling and
synchronization.
We consider two stiff helices that are modelled by rigidly connected
beads, neglecting any elastic deformations. They are driven by
constant and equal torques, and they are fixed in space by anchoring
their terminal beads in harmonic traps.
We observe that, for finite trap strength, hydrodynamic interactions
do indeed synchronize the helix rotations. The speed of phase
synchronization decreases with increasing trap stiffness. In the limit 
of infinite trap stiffness, 
the speed is zero and
the helices do not synchronize.
%
%
%
}

\maketitle


\section{Introduction}
\label{sec:introduction}

Many types of bacteria, such as certain strains of
{\it Escherichia coli} or {\it Salmonella typhimurium}, swim by
rotating flagellar filaments, which are several micrometers long and 
about 20~nm in diameter (the size of the cell body is about 1~$\mu$m)
\cite{ber73,mac87,tur00,ber03,ber04}.
The complete flagellum consists of three parts: the basal body which
is a reversible rotary motor embedded in the cell wall, the helical
filament that acts as propellor, and in-between a short flexible
coupling called the proximal hook \cite{mac87,tur00,ber03,ber04}.
The motor is powered by protons moving down an electrochemical
gradient \cite{mac87,ber03,ber04}, which generates a constant torque
independent of the dynamic load \cite{ber03,man80}. The rotation rates
for the flagella of freely moving bacteria are of the order of 100~Hz
\cite{tur00,ber03,ber04}.
The filaments are polymers with high flexural and torsional stiffness
\cite{mac87,ber03,ber04}. However, they are flexible enough to switch
between different helical forms with distinct curvature and twist
\cite{tur00,ber03,ber04}.

Typically, the filaments rotate in synchrony, i.e., the helices are 
locked in phase so that they can form bundles. As a result, the cell
is propelled at swimming speeds of about 30~$\mu$m/s
\cite{ber73,tur00,ber04}.
The process of bundling of nearby rotating ``filaments'' was 
studied in detail in macroscopic-scale experiments
\cite{mac77,kim03,kim04a}. The cell tumbles and changes its direction of
swimming randomly when one or more of the flagellar motors reverses
its direction which forces the flagellar filaments to leave the
bundle. In addition, a sequence of changes in the filament's
handedness and pitch occurs \cite{tur00,ber03,ber04,mac77b}.
Hence, the overall movement of a bacterium is the result of altering
intervals of tumbling and straight swimming. Chemotaxis steers the
bacterium by just regulating the tumbling frequency so that the net
motion heads for a more favorable food environment \cite{ber03,ber04}.

For an object with a characteristic linear dimension $a$ moving with
velocity $v$ through a Newtonian fluid, 
the ratio of inertial to viscous forces
is given by the Reynolds number $\text{Re}=av\rho/\eta$, where $\rho$
is the fluid density and $\eta$ the viscosity \cite{pur77,dho96}.
Therefore, at low Reynolds numbers ($\text{Re}\ll 1$), inertia does
not play an important role, and the thrust pushing the object forward
results solely from viscous drag.
Swimming microorganisms in water are moving at very low Reynolds
numbers \cite{pur77}. {\it E.\ coli} bacteria, e.g., have a cell body
of size $a\approx 1\mu\text{m}$ and move with velocities of the order
of $v\approx 10\mu\text{m}/s$, which yields
$\text{Re}\approx 10^{-5}$.
Thus, the locomotion of microorganisms is fundamentally different
from propulsion mechanisms in the macroscopic world (for comparison, a
dolphin moves at $\text{Re}\approx10^{7}$).

At low Reynolds numbers, the relative motion of two objects is
governed by long-range hydrodynamic interactions which, to leading
order, fall off with their inverse distance \cite{dho96}. They are
also important in biological systems. Having in mind the propulsion
mechanism of spermatozoa, Taylor modeled the hydrodynamics of two
neighboring undulating tails, and found that hydrodynamic interactions
synchronize the phases of lateral waves traveling down the tails
\cite{tay51}. Furthermore, the coordinated motion or stroke of beating
cilia (known as metachronism) is believed to be mediated by hydrodynamic
coupling \cite{lag02,lag03,gue99}. In both cases, it is observed that
the overall friction in the system is reduced by synchronization
\cite{tay51,gue99}. In analogy to these examples, it was suggested 
that hydrodynamic interactions may also play an important role in how 
flagellar filaments synchronize their rotational motion so that they
can form bundles \cite{ber73}.

In a recent paper, Kim and Powers studied hydrodynamic interactions
between two rotating helices within the framework of slender-body
theory \cite{kim04b}. The helices were considered as rigid and prevented from
translation by external forces, so that their axes were always
parallel. The key result of this work was that there is no phase
synchronization in this setup, when the two helices are driven with
the same torque.

In this paper, we consider a model which also consists of two
{\it stiff}\,  helices, thus neglecting any effects of elastic
deformations. The helices are modeled by single beads that are rigidly
connected to each other and are driven by constant and equal torques.
In contrast to Ref.~\cite{kim04b}, we ``fix'' the helices 
in space by anchoring their terminal beads in harmonic traps. This
allows for slight shifts and tilts of the helices and thus implies
some kind of {\it flexibility}, which is the major difference to
Ref.~\cite{kim04b}.
In the following, we show that the phases of the rotating helices 
do indeed synchronize in this setup, and that the state of zero phase
difference possesses lowest friction.

\begin{figure}
\includegraphics[width=0.49\columnwidth]{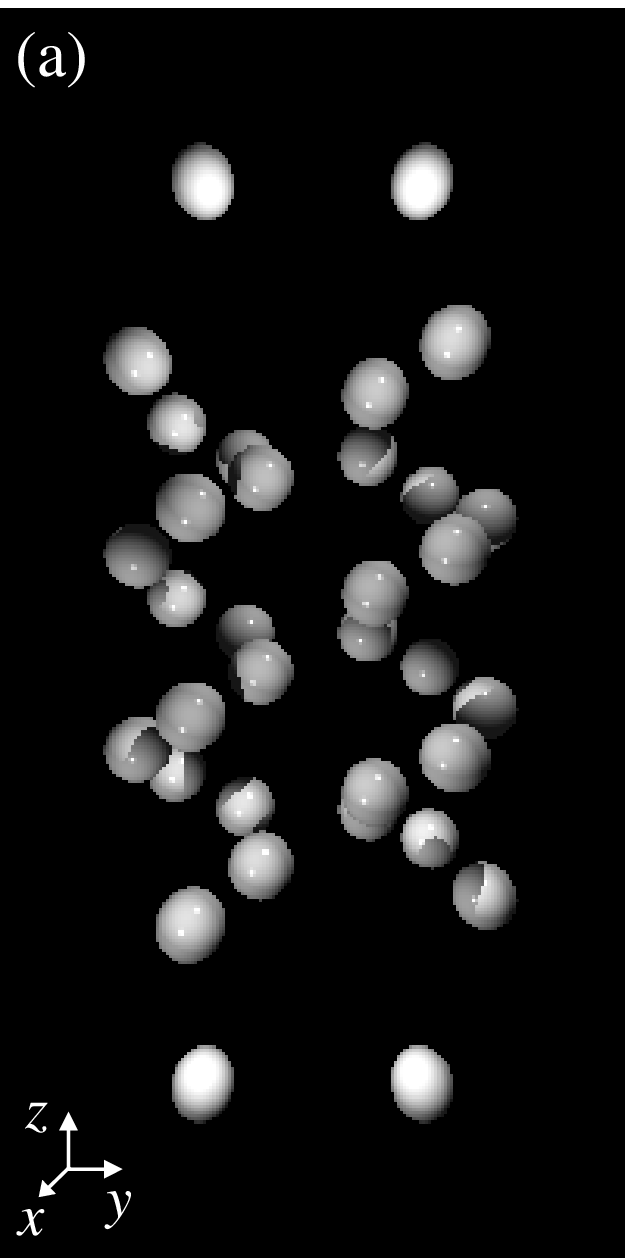}%
\hfill%
\includegraphics[width=0.49\columnwidth]{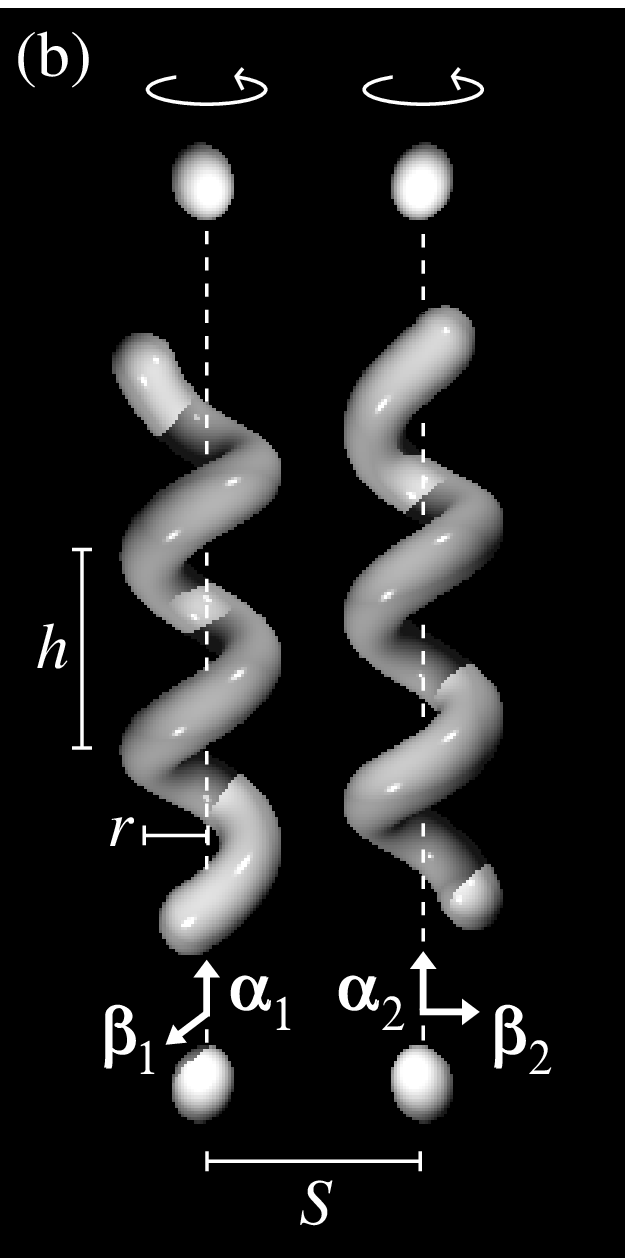}%
\caption{Visualization of the helix geometry used in the
simulations (here with a phase difference of $\pi/2$). (a) All beads
of one helix are connected ridigly with each other. (b) For the sake
of clarity, the beads are ``smeared'' out along the helix.
The top and bottom beads are anchored in harmonic traps. 
The illustrated helices are in their equilibrium positions (i.e., in
the absence of driving torques).
}
\label{fig:model}
\end{figure}

The model is introduced in detail in Sec.~\ref{sec:model}.
Then, symmetry properties of the dynamic quantities are
derived in Sec.~\ref{sec:symmetry}.
The numerical simulations of the helix dynamics are presented in
detail in Sec.~\ref{sec:synchronization}, where we analyze the data
for phase synchronization, in particular with respect to the 
anchoring strength of the terminal beads.
Finally, we conclude in Sec.~\ref{sec:conclusions} discussing the
role of the harmonic traps and the flexibility which they create.


\section{Model}
\label{sec:model}

We consider two identical helices built of equal-sized beads
[Fig.~\ref{fig:model}(a)] that are connected with each
other by (virtual) rigid bonds. Thus, the helices
cannot deform elastically. The centers of the beads are aligned along
the backbone of the helix, with equal distances between successive
beads.

\begin{figure}
\sidecaption
\includegraphics[width=0.45\columnwidth]{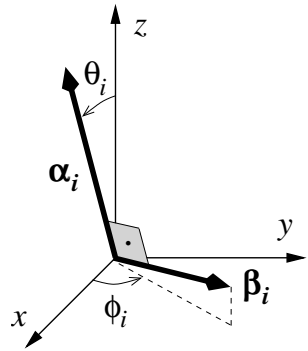}
\caption{The helix axis $\vec{\alpha}_{i}$ is tilted by the angle
$\theta_{i}$ against the $z$ direction. The phase $\phi_{i}$ measures the
angle between the $xy$-plane projection of the phase vector $\vec{\beta}_{i}$
and the $x$ axis.}
\label{fig:angles}
\end{figure}

To describe the dynamics of the helices, we introduce body-fixed
coordinate axes, given by the orthonormal vectors $\vec{\alpha}_{i}$,
$\vec{\beta}_{i}$, and $\vec{\alpha}_{i}\times\vec{\beta}_{i}$
($i=1,2$).
The axis of a helix is represented by $\vec{\alpha}_{i}$, and
the orientation of the perpendicular vector $\vec{\beta}_{i}$ shall
describe the phase of the helix, i.e., the rotation about its own axis
[Fig.~\ref{fig:model}(b)].
We define the phase angles $\phi_{i}$ by the projection of
$\vec{\beta}_{i}$ into the $xy$ plane (Fig.~\ref{fig:angles}).
The angle between $\vec{\alpha}_{i}$ and the $z$ axis is
the tilt angle $\theta_{i}$.

The centers of mass of the helices are denoted by $\vec{x}_{i}$. The
positions of the individual beads are then given by 
\begin{equation}
\bar{\vec{x}}_{i}^{\nu}=\vec{x}_{i}
+\xi_{1}^{\nu}\vec{\alpha}_{i}
+\xi_{2}^{\nu}\vec{\beta}_{i}
+\xi_{3}^{\nu}\vec{\alpha}_{i}\times\vec{\beta}_{i}
\end{equation}
with the internal coordinates
\begin{equation}
\label{eq:helix_coord}
\begin{array}{c}
\displaystyle
\xi_{1}^{\nu}=\frac{h}{m}\left(\nu-\frac{nm-1}{2}\right) \, ,
\\[6mm]
\displaystyle
\xi_{2}^{\nu}=r\cos\frac{2\pi}{m}\,\nu \, ,
\quad
\displaystyle
\xi_{3}^{\nu}=r\sin\frac{2\pi}{m}\,\nu \, ,
\end{array}
\end{equation}
where $r$ is the radius of the helix and $h$ its pitch.
The bead index $\nu$ runs from $0$ to $nm-1$ for each helix, with $m$
being the number of beads per winding and $n$ the number of
windings.

\begin{figure*}
\sidecaption
\includegraphics[width=0.65\columnwidth]{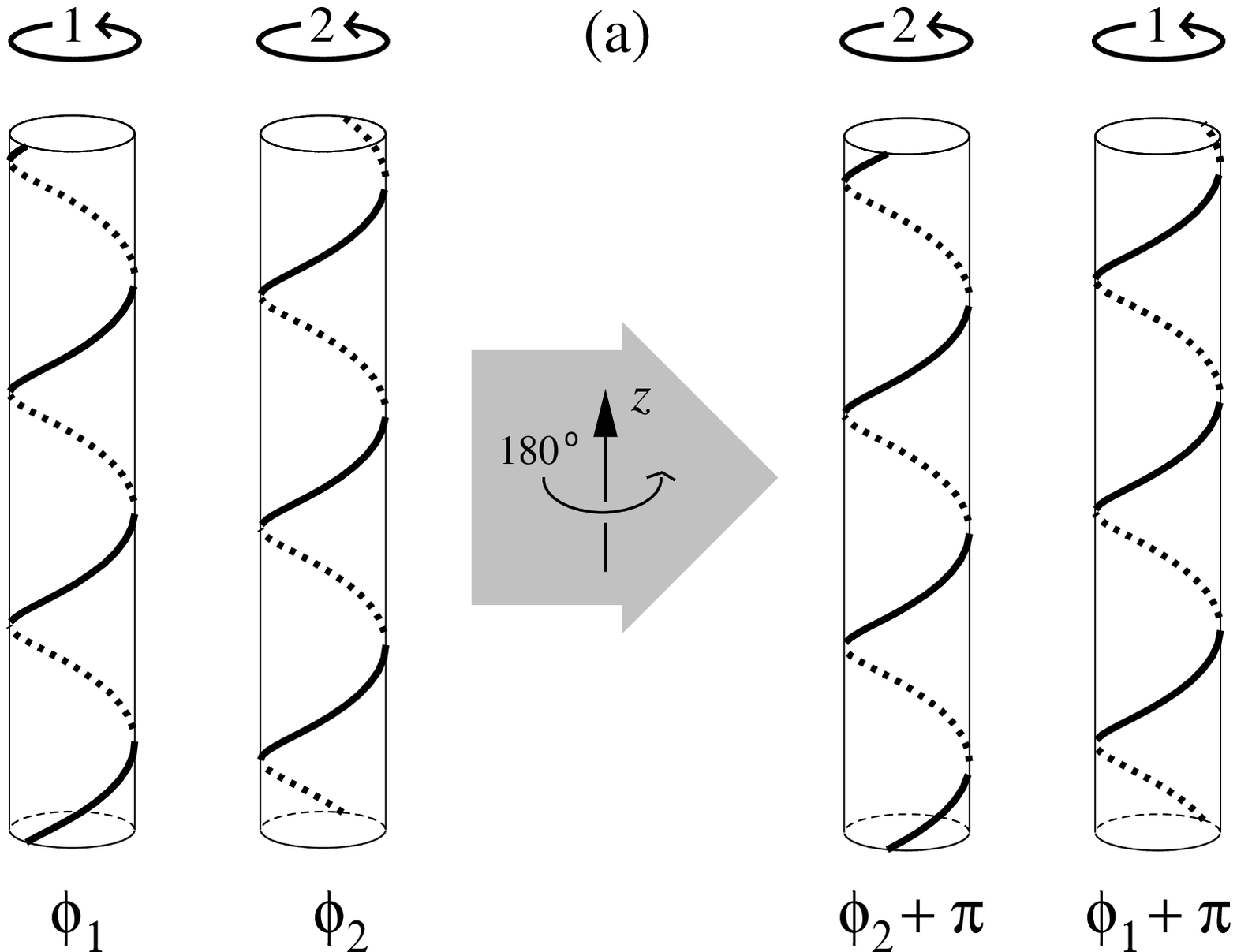}%
\hspace{1.0cm}%
\includegraphics[width=0.65\columnwidth]{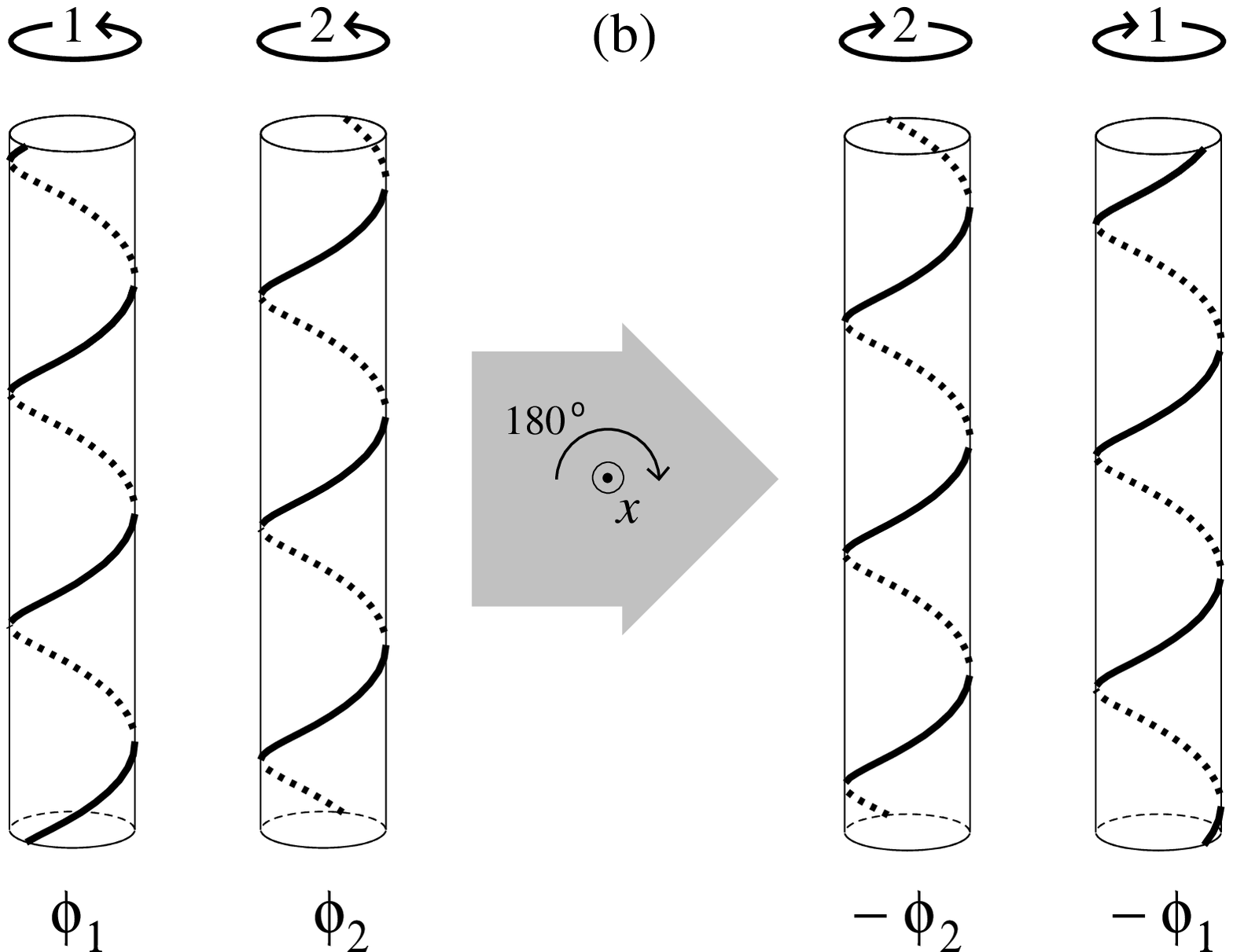}%
\hspace{5mm}%
\caption{Rotation of the two-helix system by $180^{\circ}$ about the
$z$ axis (a) and about the $x$ axis (b).
The circular arrows (indexed with $i=1,2$) on top of the helices
denote that the respective helix is driven with torque $T_{i}$ and
rotates with velocity $\omega_{i}$ in the indicated direction.
The tubes are drawn as guide to the eye.}
\label{fig:symmetry}
\end{figure*}

The helices are driven by constant and equal torques
that are always parallel to the respective helix axis, i.e., the
torques are given by $D\vec{\alpha}_{i}$ with a fixed
parameter $D$.
Note that the assumption of a constant torque agrees with experimental
studies of real flagellar motors \cite{ber03,man80}, as we have
already mentioned in our introductory remarks.
To ``fix'' the helices in space, we attach single beads 
at the top and bottom end of each helix axis (Fig.~\ref{fig:model})
and anchor them in harmonic traps with equal force constants $K$. In
equilibrium, both helix axes are parallel, and their center-to-center
distance is $S$. If one of the anchoring beads is displaced by
$\Delta\bar{\vec{x}}_{i}^{\sigma}$ (where the index $\sigma$ refers to
``top'' or ``bottom'') relative to the center of the respective
harmonic trap, the restoring {\it single-particle} force is
\begin{equation}
\bar{\vec{F}}_{i}^{\sigma}=-K\Delta\bar{\vec{x}}_{i}^{\sigma} \, .
\end{equation}
Finally, the total {\it center-of-mass} forces and torques
acting on the rigid helices are 
\begin{equation}
\label{eq:forces+torques}
\begin{array}{l}
\vec{F}_{i}=\sum\limits_{\sigma}\bar{\vec{F}}_{i}^{\sigma}
\quad (\text{with}\ \sigma=\text{top, bottom}) \, , \\[4mm]
\vec{T}_{i}= D\vec{\alpha}_{i}
+\sum\limits_{\sigma}(\bar{\vec{x}}_{i}^{\sigma}-\vec{x}_{i})
\times\bar{\vec{F}}_{i}^{\sigma} \, .
\end{array}
\end{equation}

In the regime of low Reynolds numbers, the flow of an incompressible
fluid with viscosity $\eta$ obeys the quasi-static Stokes or creeping
flow equations $\eta\nabla^{2}\vec{u}-\nabla p=\vec{0}$ and
$\nabla\cdot\vec{u}=0$ \cite{dho96,bre63/64}, where
$\vec{u}$ is the flow field and $p$ the hydrodynamic pressure. We
assume the flow to vanish at infinity and impose stick boundary
conditions on the surfaces of all particles suspended in the fluid.
The resulting flow field then couples the motion of the particles
to each other.
Due to the linearity of the Stokes equations, 
their translational and rotational
velocities, $\vec{v}_{i}$ and $\vec{\omega}_{i}$, depend linearly on
all external forces and torques, $\vec{F}_{j}$ and $\vec{T}_{j}$
\cite{dho96,bre63/64}:
\begin{equation}
\label{eq:hydro_int_gen}
\begin{array}{l}
\vec{v}_{i}
=\sum\limits_{j}\boldsymbol{\mu}_{ij}^{\text{tt}}\vec{F}_{j}
+\sum\limits_{j}\boldsymbol{\mu}_{ij}^{\text{tr}}\vec{T}_{j}
\, , \\[4mm]
\vec{\omega}_{i}
=\sum\limits_{j}\boldsymbol{\mu}_{ij}^{\text{rt}}\vec{F}_{j}
+\sum\limits_{j}\boldsymbol{\mu}_{ij}^{\text{rr}}\vec{T}_{j}
\, .
\end{array}
\end{equation}
Each of the mobilities $\boldsymbol{\mu}_{ij}^{\text{tt}}$,
$\boldsymbol{\mu}_{ij}^{\text{tr}}$,
$\boldsymbol{\mu}_{ij}^{\text{rt}}$, and
$\boldsymbol{\mu}_{ij}^{\text{rr}}$
is a $3\times 3$ tensor, 
which couples the translations (superscript t) and rotations
(superscript r) of particles $i$ and $j$.
They depend on the current spatial configuration of all suspended
particles. Since this dependence is highly nonlinear, they 
have to be calculated numerically.

In our simulations, we use the numerical library {\sc hydrolib}
\cite{hin95} which yields the full set of mobility tensors for a given
configuration of equal-sized spherical particles (based on the
multipole expansion method). It implicitly accounts for (virtual)
rigid bonds that keep the relative positions of the single beads in a
rigid cluster fixed. Thus, {\sc hydrolib} calculates an effective
mobility matrix for the coupled center-of-mass translations and
rotations, i.e., the indices $i$ and $j$ in
Eq.~(\ref{eq:hydro_int_gen}) now refer to rigid clusters instead of
individual beads (for details, see Ref.~\cite{hin95}).

Therefore, with the forces and torques given in
Eq.~(\ref{eq:forces+torques}), we directly obtain the linear and
angular velocities of the helices. The translational motion
of the centers of mass is then governed by
\begin{equation}
\label{eq:x'=v}
\dot{\vec{x}}_{i}=\vec{v}_{i} \, ,
\end{equation}
where the dot means time derivative.
The rotational motion of the helix axes $\vec{\alpha}_{i}$ and the
phase vectors $\vec{\beta}_{i}$ follows from
\begin{equation}
\label{eq:alpha'=omega*alpha}
\begin{array}{l}
\dot{\vec{\alpha}}_{i}=\vec{\omega}_{i}\times\vec{\alpha}_{i} \, ,
\\[2mm]
\dot{\vec{\beta}}_{i}=\vec{\omega}_{i}\times\vec{\beta}_{i} \, .
\end{array}
\end{equation}
We integrate these equations in time by applying a second-order
Runge-Kutta scheme (also known as Heun algorithm) \cite{kloe99}. Note
that the mobility matrices have to be evaluated at each time step
since the positions and orientations of the helices change.

While the trap constant $K$ was varied to study 
the influence of the anchoring strength on the helix dynamics,
the driving torque $D$ was
kept fixed since it merely sets the time scale (given by the
rotational frequency $\omega_{0}$ of an isolated helix).
The time steps of the numerical integration where chosen to correspond
to roughly 1/360th of a revolution of a single helix.

The geometry of the two helices is shown in
Fig.~\ref{fig:model}. Their backbones have a radius of $r=2.0a$ and a
pitch of $h=6.0a$, where $a$ is the bead radius.
The number of windings is $n=3$, and the number of beads per winding
is $m=5$. The distance between the anchoring beads and the helix is
the same as the pitch $h$. The equilibrium separation of the helices,
i.e., the distance of the upper/lower anchoring traps, is $S=7.0a$.
Note that the calculation of the mobility matrix is the most time
consuming part in the simulations. Therefore, we had to restrict the
number of beads in one helix. Furthermore, we will only present
results for the set of parameters just introduced and concentrate on
the essential variable, namely the trap stiffness $K$.


\section{Symmetry considerations}
\label{sec:symmetry}

Consider for the moment two helices whose axes are completely fixed in
space, i.e., translation and tilt are prevented by appropriate forces
and torques. In this case, the only remaining degrees of freedom are
rotations about the axes of the helices. They are described by the
phase angles $\phi_{i}$ and the angular velocities 
$\dot{\phi}_{i}=\omega_{i}$. 
We introduce the phase difference $\chi=\phi_{2}-\phi_{1}$
as the phase of the right helix relative to the left helix, when
viewed as in Fig.~\ref{fig:symmetry}(a,b) (left part).
According to Eq.~(\ref{eq:hydro_int_gen}), the rotational velocities
$\omega_{i}$ are functions of the phase angles $\phi_{i}$ since the
mobility tensors depend on the spatial configuration.
As the helices are driven by the same torques $T_{i}=D$ about their
axes, the synchronization rate is given by 
\begin{equation}
\label{eq:chi'=mu*D}
\dot\chi=\omega_{2}(\phi_{1},\phi_{2})-\omega_{1}(\phi_{1},\phi_{2})
=\mu(\phi_{1},\phi_{2})D \, ,
\end{equation}
where the effective mobility $\mu$ is $2\pi$-periodic in $\phi_{i}$.
We choose this careful definition because we now want to derive
symmetry properties of $\mu$.

We use the fact that the dynamics of the two-helix system must not
change under arbitrary rotations of the whole geometry since the
surrounding fluid is iso\-tropic. By applying the two operations
illustrated in Fig.~\ref{fig:symmetry}, we create new configurations
with left and right helices whose known dynamics we use to infer
properties of the mobility~$\mu$.

In the first case, we rotate the two-helix system
by $180^{\circ}$ about the $z$ axis, as illustrated in
Fig.~\ref{fig:symmetry}(a). The velocities of the left and right
helix are exchanged, i.e., $\omega_{1}\leftrightarrow\omega_{2}$ and
thus $\dot{\chi}\to-\dot{\chi}$. 
On the other hand, the phase angles of the new left and right helix
are, respectively, $\phi_{2}+\pi$ and $\phi_{1}+\pi$. Combining
both statements, Eq.~(\ref{eq:chi'=mu*D}) yields
\begin{equation}
\label{eq:symm_z}
\mu(\phi_{2}+\pi,\phi_{1}+\pi)=-\mu(\phi_{1},\phi_{2})
\, .
\end{equation}
If the phases of the helices differ by $\pi$ ($\phi_{2}=\phi_{1}+\pi$),
one therefore obtains 
$\mu(\phi_{1},\phi_{1}+\pi)=-\mu(\phi_{1},\phi_{1}+\pi)=0$ or
\begin{equation}
\label{eq:dotchi(chi=pi)}
\dot{\chi}=0 \quad \text{for} \quad \chi=\pi \, ,
\end{equation}
i.e., the synchronization speed vanishes for any $\phi_{1}$ whenever
$\phi_{2}-\phi_{1}=\pi$.

Let us now rotate the two-helix system by $180^{\circ}$ about the $x$
axis [Fig.~\ref{fig:symmetry}(b)].
Then the velocities of the left and right helix are exchanged and
reversed, i.e., $\omega_{1}\leftrightarrow -\omega_{2}$, and the
synchronization speed
$\dot{\chi}=-\omega_{1}-(-\omega_{2})=\omega_{2}-\omega_{1}$ 
stays the same. On the other hand, the angles transform as
$\phi_{1}\leftrightarrow-\phi_{2}$, and the driving torques are
reversed, i.e., $D\to -D$. 
Again, combining both statements, Eq.~(\ref{eq:chi'=mu*D}) yields
$\mu(-\phi_{2},-\phi_{1})(-D)=\mu(\phi_{1},\phi_{2})D$,
and thus
\begin{equation}
\label{eq:symm_x}
\mu(-\phi_{2},-\phi_{1})=-\mu(\phi_{1},\phi_{2}).
\end{equation}
For helices of infinite length, the dynamics can only depend on the
phase difference $\chi$ and not on the single phases $\phi_{i}$. This
is obvious since a phase shift of both helices is equivalent to a
translation along the helix axes, which does not change the
dynamics. Hence, Eq.~(\ref{eq:symm_x}) reads
$\mu(\chi)=-\mu(\chi)=0$, i.e., for parallel helices of
infinite length, the synchronization rate $\dot{\chi}=\mu(\chi)D$
vanishes for any phase difference $\chi$, and therefore, they do not
synchronize towards $\chi = 0$ \cite{foot:kin_rev}.


\section{Synchronization}
\label{sec:synchronization}

We now study the rotational dynamics of two helices whose terminal
beads are anchored in harmonic traps of {\it finite} strength, as
introduced in Sec.~\ref{sec:model}. Thus, the helices can be shifted
and tilted, and their axes undergo a precession-like motion while
each helix itself is rotating about its respective axis.
The orientations of the helices in space are described by the vectors
$\vec{\alpha}_{i}$ and $\vec{\beta}_{i}$ [see Fig.~\ref{fig:model}(b)]
and the corresponding angles $\theta_{i}$ and $\phi_{i}$, as defined
in Fig.~\ref{fig:angles}.

\begin{figure}
\includegraphics[width=\columnwidth]{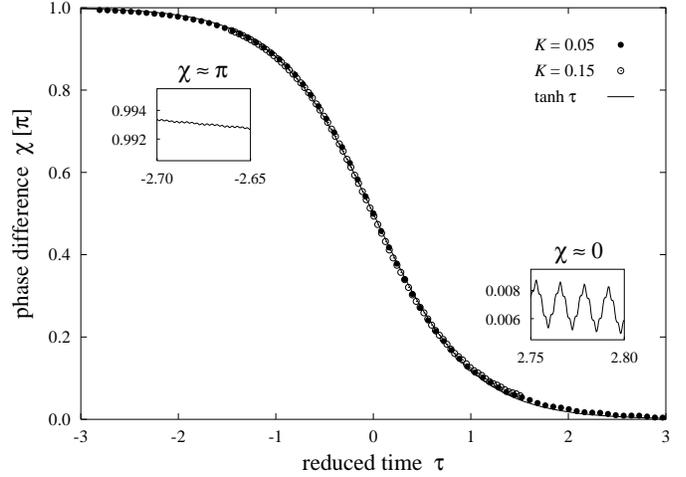}
\caption{Synchronization of the helix rotations. The phase difference
of the two helices tends towards zero, starting from $\chi$
slightly smaller than $\pi$.
The symbols are simulation data at two different trap strengths
(values of $K$ in units of $D/a$). For clarity, not every data point
is plotted. The solid line shows the master curve of
Eq.~(\ref{eq:phase_sync}). The insets enlarge the small oscillations
at $\chi\approx 0$ and $\pi$ (here for the case $K=0.05\,D/a$, but the
amplitudes do not depend strongly on $K$). Note that the scaling of
the two insets is the same.}
\label{fig:phase_sync}
\end{figure}

Figure~\ref{fig:phase_sync} shows the phase difference 
$\chi=\phi_{2}-\phi_{1}$  for two trap stiffnesses $K$
as a function of a reduced time $\tau(K)$, to be defined below.
Starting with $\chi$ slightly smaller than $\pi$, the phase
difference decreases continuously (with steepest slope at
$\chi=\pi/2$) and finally approaches zero, i.e., the two helices do
indeed synchronize their phases. 
The simulations reveal that the dynamics does not depend significantly
on the phases $\phi_{i}$ themselves, but is predominantly determined
by the phase {\it difference} $\chi$. 
Note that this feature may be expected since the dynamics of parallel
helices of infinite length can only depend on $\chi$ as explained in 
Sec.~\ref{sec:symmetry}. To be concrete, we observe that the
rotational velocities $\dot{\phi}_{i}$ undergo oscillations of only
about 1\% around a mean value during one rotational period.
Their amplitude does not depend on the trap stiffness $K$, i.e., the
oscillations originate from the slight dependence on the phases
themselves (and not from the precession of the axes).
Therefore, starting with different values for $\phi_{i}$ but the same
value of $\chi$ yields the same curve (except for differences in the
small oscillations illustrated in the insets of
Fig.~\ref{fig:phase_sync}).

\begin{figure}
\includegraphics[width=\columnwidth]{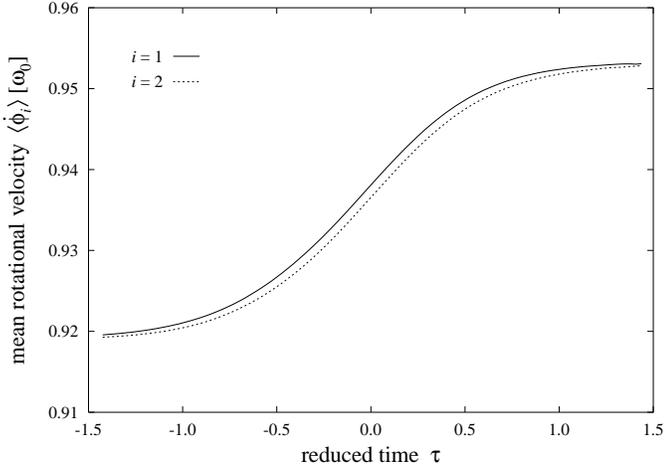}
\caption{Mean rotational velocities, averaged over one revolution (in
units of the rotational frequency $\omega_{0}$ of an isolated helix).
The oscillations about the mean value are of the order of a few
percent, they decrease slightly during the synchronization
process. The example shown is for trap strength $K=0.15\,D/a$.}
\label{fig:rot_speed}
\end{figure}

The mean rotational velocities, averaged over one
rotational period, increase during the synchronization process from
about $0.92\omega_{0}$ at $\chi\approx\pi$ to about $0.95\omega_{0}$ at
$\chi\approx 0$ (Fig.~\ref{fig:rot_speed}).
Thus, the hydrodynamic drag acting on the helices is minimized during
phase synchronization (see also Ref.~\cite{kim04b}). Since the torques
are constant, the dissipation rate
$\sum_{i}\vec{T}_{i}\cdot\vec{\omega}_{i}$ is maximized.
This observations agrees with the interesting fact that the Stokes equations
can be derived from a variational principle where one searches for
an extremum of the dissipated energy $\int \sigma_{ij} A_{ij} d^{3}r$
($\sigma_{ij}$ is the stress tensor and $A_{ij}$ the symmetrized
velocity gradient) under the constraint that the fluid is incompressible
\cite{lam72,fin72}.
The pressure enters via the Lagrange parameter associated with the
constraint.

In Sec.~\ref{sec:symmetry}, we showed for fixed parallel helices, 
based on pure symmetry arguments, that their synchronization rate 
vanishes for a phase difference of $\chi=\pi$ 
[see Eq.~(\ref{eq:dotchi(chi=pi)})]. In that case, the two-helix 
configuration is symmetric with respect to a rotation by $180^{\circ}$ 
about the $z$ axis [see Fig.~\ref{fig:symmetry}(a)]. Our reasoning of
Sec.~\ref{sec:symmetry} can be extended to the case of non-parallel
helix axes, as long as the same symmetry is preserved. 
However, $\chi=\pi$ does not correspond to a stable state. Starting 
with $\chi$ marginally smaller than $\pi$, the system tends
towards phase difference zero.
The simulation with $K=0.05$ in Fig.~\ref{fig:phase_sync} was launched, 
e.g., at $\chi=0.994\pi$ with both helices in equilibrium
position and orientation, as shown in Fig.~\ref{fig:model}.

\begin{figure}
\includegraphics[width=\columnwidth]{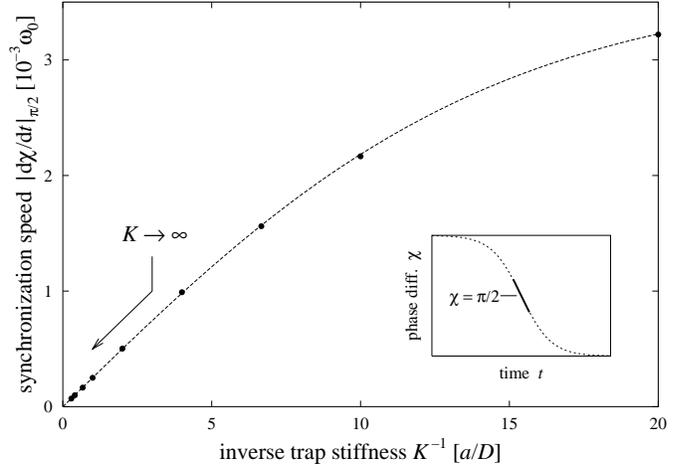}
\caption{Synchronization speed (taken at a relative phase
$\chi\approx\pi/2$, as illustrated in the inset) as a function of 
inverse trap strength $K^{-1}$. 
The frequency scale $\omega_{0}$ for the synchronization speed
is the angular velocity of an isolated helix. 
The symbols indicate values extracted from
simulations at different $K$. The dashed line is an empirical fit (see
text).}
\label{fig:sync_speed}
\end{figure}

On the other hand, the synchronized state
$\chi=0$ is stable against small 
perturbations since configurations with $\chi$ between 0 and $-\pi$
synchronize towards zero phase difference, too. This was checked by 
simulations, but can also be derived from Eq.~(\ref{eq:symm_z}).
The corresponding rotation of Fig.~\ref{fig:symmetry}(a) creates
new left and right helices with a change in sign for $\chi$ and
$\dot{\chi}$ relative to the original helices which explains our
statement. Furthermore, starting a simulation with exactly 
$\chi=0$, the helices remain synchronized on average 
(i.e., $\langle\chi\rangle=0$), but there are still small oscillations 
as illustrated in the lower right inset in Fig.~\ref{fig:phase_sync}
for the case where both helices started in equilibrium position and 
orientation.

Averaging over small oscillations, we find that the resulting smoothed
curves for the phase difference $\chi$ obey an empirical law of the
form
\begin{equation}
\label{eq:phase_sync}
\chi(\tau)=\frac{\pi}{2}(1-\tanh\tau) \, ,
\end{equation}
where
\begin{equation}
\label{eq:red_time}
\tau(K)=\frac{2}{\pi}\,(t-t_{\pi/2})
\left|\frac{\text{d}\chi}{\text{d}t}\right|_{t=t_{\pi/2}}
\end{equation}
is the reduced time, already mentioned above, and $t_{\pi/2}$ denotes 
the time where $\chi=\pi/2$, i.e., the location of the inflection
point. Its slope $|\text{d}\chi/\text{d}t|_{t=t_{\pi/2}}$ depends on
the trap stiffness $K$ and so does $\tau$.
As Fig.~\ref{fig:phase_sync} strikingly reveals, this
law works very well. By plotting the phase difference $\chi$
versus the reduced time $\tau(K)$, the curves collapse on the
master curve given by Eq.~(\ref{eq:phase_sync}).
Since the dynamics at low Reynolds numbers is completely overdamped,
we expect this law to follow from a differential equation which is of
first order in time. Taking the first derivative of
Eq.~(\ref{eq:phase_sync}) with respect to $\tau$, we find that
$\chi(\tau)$ obeys the nonlinear equation 
$\dot{\chi}(\tau)=(2/\pi)\chi(\tau)[\pi-\chi(\tau)]$, known as the
Verhulst equation and originally proposed to model the development of
a breeding population \cite{arf95}.
However, it is not clear how to derive this equation from first
principles in our case.

\begin{figure}
\includegraphics[width=\columnwidth]{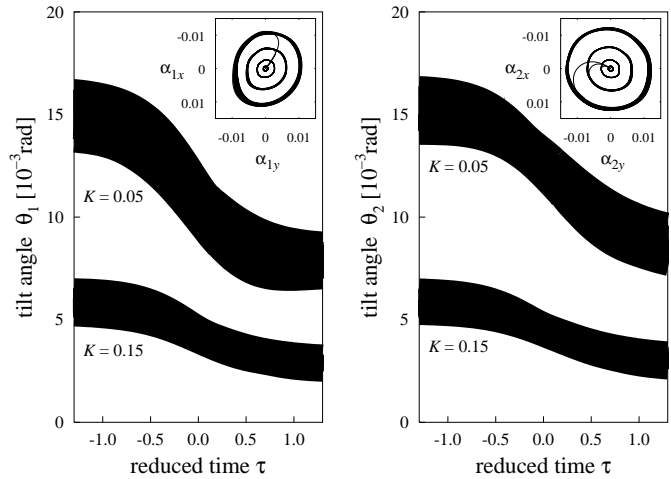}
\caption{Tilt of the left and right helix axis for $K=0.05$ and $0.15$ as 
a function of reduced time $\tau$.
The fast periodic oscillations of the tilt angles $\theta_{i}$ yield the 
black bands since they cannot be resolved on the time scale used
here. The insets visualize the precession of the helix axes by showing
the tip of the vectors $\vec{\alpha}_{i}$ (at $\chi\approx\pi/2$). 
From the outer to the inner ``circular'' orbit, the trap stiffness
assumes the values $K=0.05$, 0.1, 0.15, and 1.0 (in units of $D/a$).}
\label{fig:tilt}
\end{figure}

An important result is that the speed of the synchronization process 
decreases with increasing trap stiffness $K$. The values plotted in 
Fig.~\ref{fig:sync_speed} for different $K^{-1}$
are the slopes $|\text{d}\chi/\text{d}t|_{t=t_{\pi/2}}$ extracted from
simulation data at the inflection point with a relative phase of
$\chi\approx\pi/2$. 
The curve in Fig.~\ref{fig:sync_speed} can be extrapolated by the analytic 
form $c_{1}\tanh c_{2}K^{-1}$ (dashed curve), where the fit parameters
assume the values $c_{1}=3.67\cdot10^{-3}\omega_{0}$ and
$c_{2}=0.0685\,D/a$.
In the limit of infinite trap strength, i.e., for $K^{-1}\to 0$,
the synchronization speed clearly tends towards zero, i.e., an
infinitely strong anchoring of the helix axes does not allow for 
phase synchronization.

In Fig.~\ref{fig:tilt}, we illustrate how the tilt angles $\theta_{i}$
(for their definition, see Fig.~\ref{fig:angles}) vary during the
synchronization process. The mean tilt angle as well as the amplitude
of its periodic oscillations decrease when the phase difference
approaches zero. Obviously, the dynamics of the helices depends on the
stiffness of the harmonic anchoring of the top and bottom terminal
beads. In a weaker trap, the tilt of the helix axes out of equilibrium
is more pronounced compared to a stronger trap.
The insets in Fig.~\ref{fig:tilt} track the precession-like motions of
the helix axes. The stronger the trap, the smaller the radius of the
``orbit'' or the tilt angle. 
[Since the simulations were started with both axes aligned along their
equilibrium direction, the trajectories $(\alpha_{ix},\alpha_{iy})$
first move radially away from the origin and then enter the
``precession orbit''.]

\begin{figure}
\includegraphics[width=\columnwidth]{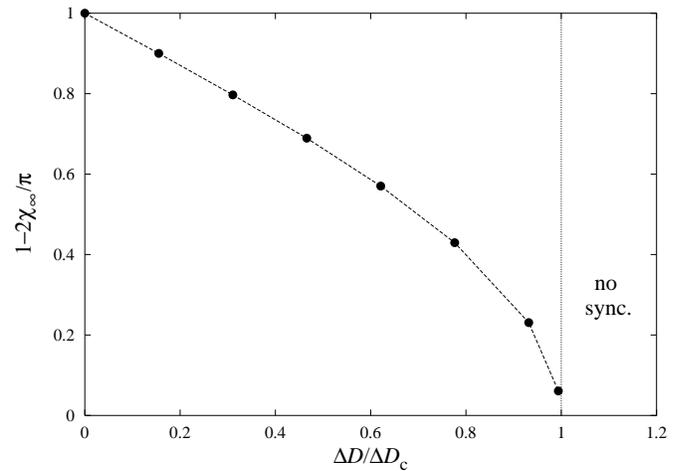}
\caption{Phase difference $\chi_{\infty}$ (plotted as
$1-\chi_{\infty}/\frac{\pi}{2}$) of the synchronized state as 
a function of the torque difference $\Delta D$. The phase difference
increases from $\chi_{\infty}=0$ to $\pi/2$ (or
\mbox{$1-\chi_{\infty}/\frac{\pi}{2}$} decreases from 1 to 0) with
increasing torque difference. For $\Delta D>\Delta D_{\text{c}}$,
there is no synchronization.}
\label{fig:sync_phase_diff}
\end{figure}

In real flagellar motors, the torques on the two helical filaments are
not exactly the same. To test whether the phenomenon of synchronization
still occurs, we now consider slightly different 
driving torques for the two helices. The first helix is still driven
with torque $D$, while the second helix is driven with
$D+\Delta D$ ($\Delta D>0$). We observe that, for torque differences
$\Delta D$ below a critical value $\Delta D_{\text{c}}$, 
the two helices indeed synchronize towards a phase difference 
$\chi_{\infty}=\chi(t\to\infty)$, which, in general, is not zero.
The results are shown in Fig.~\ref{fig:sync_phase_diff}, 
where we plot $1-\chi_{\infty}/\frac{\pi}{2}$ as a function
of $\Delta D/\Delta D_{\text{c}}$.
With increasing $\Delta D$, the phase lag $\chi_{\infty}$ increases
from zero to $\pi/2$. For $\Delta D>\Delta D_{\text{c}}$, there is no
synchronization, and the phase difference grows continuously.
Note that the reduced critical torque difference
$\Delta D_{\text{c}}/D=3.22\cdot 10^{-3}$ corresponds to the reduced
frequency difference
$\dot{\chi}/\omega_{0}=(\omega_{2}-\omega_{1})/\omega_{0}$ 
observed at $\chi=\pi/2$ for equal torques ($\Delta D=0$).
At the critical torque difference, the helices synchronize towards 
a phase lag of $\chi=\pi/2$. This means that $\Delta D_{\text{c}}$
just compensates the difference in the effective
mobilities of the two helices, which is largest for $\chi=\pi/2$;
thus the helices rotate with the same speed.

At the end, we mention that all results presented here refer
to helices whose rotational direction is given in
Fig.\ \ref{fig:model}(b).
Reversing the direction of rotation does not change the dynamics of the 
two-helix system since this can also be achieved by the operation shown 
in Fig.~\ref{fig:symmetry}(b) that does not change the synchronization
speed.


\section{Conclusions}
\label{sec:conclusions}

We have reported that two rigid helices whose terminal beads
are anchored in harmonic traps and which are driven by equal torques 
synchronize to zero phase difference. 
The effect is robust, i.e., if the torques are unequal, the helices
synchronize to a non-zero phase lag below a critical torque difference.
This agrees with observations in Ref.~\cite{kim03} where the helices
do not bundle if the motor speeds are sufficiently different.
Increasing the stiffness of the anchoring traps, decreases the
synchronization rate. We attribute this to the jiggling motion of the
two helix axes which is more and more restrained.

In the limit of infinite trap strength, our results are consistent
with recent work based on slender-body theory for two rigid helices
\cite{kim04b}. If the helices are prevented from translation and their
axes are always kept parallel, then there is no synchronization
possible. Therefore, we conclude that the additional degree of freedom
due to the finite anchoring of the helix axis, i.e., the jiggling motion,
is essential to enable {\it phase synchronization} in our model.

At a first glance, our model might appear too artificial for describing
the hydrodynamic coupling of flagella. However, our
results clearly indicate that some kind of {\it flexibility} is essential
to allow for phase synchronization. In reality, this flexibility might
have its origin in elastic deformations of the rotating flagella.
Therefore, the next step is to make the flexural and torsional
stiffness of the helices in our model finite.

The helix used in our numerical investigation with radius $r=2a$ and pitch
$h=6a$ is far from the values of a real bacterial filament with $r=20a$
and $h=200a$ where $a$ is now half the filament diameter. We have
tried different paths in the parameter space $(r,h)$ to connect both
cases, i.e, our ``fat'' helix and the real slender helix. We observe that
making the helix more slender decreases the synchronization speed in 
units of $\omega_{0}$. This makes sense since the induced flow from
the rotation of slender helices is smaller. To reduce computer time,
we extrapolated the synchronization speeds for different paths in the
$(r,h)$ space towards the real helical flagellum and found that the
speed is reduced by a factor of 60 to 70 compared to the results
reported in this article. At a first glance this seems to
be a discouraging result. However, from preliminary results of helices
with finite bending and torsional flexibility, we know already that 
these factors considerably increase the synchronization speed so that
it will be of biological relevance. 

Furthermore, we made a comparison
of the resistance matrix of a real flagellum modeled by a sequence of 
spheres with resistive force theory as summarized, e.g., in 
Ref.~\cite{joh79}. We found that for motion and rotation along 
the helix axis, the single matrix elements differ by less than a factor 
of two. This convinces us that modelling a flagellum with the method
presented in this article is appropriate. For our fat helix, however,
the deviations are larger since the filament is relatively thick
compared to radius and pitch of the helix. So the conditions for the
validity of resistive force theory are not satisfied so well.

We also checked whether it is important if the helices are forced
to stay at their position or if they are allowed to
propel themselves. This was done by letting the helices move along 
the $z$ axis but still keeping them in harmonic traps along the 
$x$ and $y$ direction.
However, we did not see a significant difference in the dynamics 
of the synchronization process compared to the results presented 
in this work.

As a final remark, we point out that the following general mechanism 
may exist in systems of low Reynolds numbers: Synchronizing the motions of 
some objects via hydrodynamic interactions needs some kind of ``flexibility''.
If the motions of the objects are constrained too much, synchronization 
cannot occur. We have observed a similar behavior for particles circling
in a toroidal harmonic trap and driven by a constant tangential force
\cite{rei04}. After some transition regime, the particles reach
a synchronized state where they perform a periodic limit cycle.
We observe that for increasing trap stiffness, i.e., for decreasing
oscillations along the radial direction, the time to reach this limit
cycle increases.


\begin{acknowledgement}

This work was supported by the Deutsche Forschungsgemeinschaft through
Sonderforschungsbereich Transregio 6 ``Physics of colloidal
dispersions in external fields''.
H.S. acknowledges financial support from the Deutsche
Forschungsgemeinschaft by Grant No. Sta 352/5-1.

\end{acknowledgement}



\begin{thebibliography}{99}

\bibitem{ber73}
H.C.\ Berg, Nature {\bf 245}, 380 (1973).

\bibitem{mac87}
R.M.\ Macnab, in {\it Escherichia coli and Salmonella
typhi\-murium}, edited by F.C.\ Neidhardt (Am.\ Soc. Microbiol.,
Washington DC, 1987), p.\ 70.

\bibitem{tur00}
L.\ Turner, W.S.\ Ryu, H.C.\ Berg, J.\ Bacteriol.\ {\bf 182}, 2793
(2000). Movies of swimming {\it E.\ coli}\, can be found at
http:/\hspace{-2pt}/\linebreak
webmac.rowland.org/labs/bacteria/movies\_ecoli.html.

\bibitem{ber03}
H.C.\ Berg, Annu.\ Rev.\ Biochem.\ {\bf 72}, 19 (2003).

\bibitem{ber04}
H.C.\ Berg, {\it E.\ coli in Motion} (Springer, New York, 2004).

\bibitem{man80}
M.D.\ Manson, P.M.\ Tedesco, H.C.\ Berg, J.\ Mol.\ Biol.\ {\bf 138},
541 (1980).

\bibitem{mac77}
R.M.\ Macnab, Proc.\ Natl.\ Acad.\ Sci.\ USA {\bf 74}, 221 (1977).

\bibitem{kim03}
M.J.\ Kim, J.C.\ Bird, A.J.\ Van Parys, K.S.\ Breuer, T.R.\ Powers,
Proc.\ Natl.\ Acad.\ Sci.\ USA {\bf 100}, 15481 (2003). A movie of the
bundling sequence of two rotating (macroscopic) helices can be found
at http:/\hspace{-2pt}/www.pnas.org/cgi/\linebreak
content/full/2633596100/DC1.

\bibitem{kim04a}
M.J.\ Kim, M.J.\ Kim, J.C.\ Bird, J.\ Park, T.R.\ Powers, K.S.\
Breuer, Exp.\ Fluids {\bf 37}, 782 (2004).

\bibitem{mac77b}
R.M.\ Macnab, M.K.\ Ornston, J.\ Mol.\ Biol.\ {\bf 112}, 1 (1977).

\bibitem{pur77}
E.M.\ Purcell, Am.\ J.\ Phys.\ {\bf 45}, 3 (1977).

\bibitem{dho96}
J.K.G.\ Dhont, {\it An Introduction to Dynamics of Colloids}
(Elsevier, Amsterdam, 1996).

\bibitem{tay51}
G.\ Taylor, Proc.\ R.\ Soc.\ Lond.\ A {\bf 209}, 447 (1951).

\bibitem{lag02}
M.C.\ Lagomarsino, B.\ Bassetti, P.\ Jona, Eur.\ Phys.\ J.\ B
{\bf 26}, 81 (2002).

\bibitem{lag03}
M.C.\ Lagomarsino, P.\ Jona, B.\ Bassetti, Phys.\ Rev.\ E {\bf 68},
021908 (2003).

\bibitem{gue99}
S.\ Gueron, K.\ Levit-Gurevich, Proc.\ Natl.\ Acad.\ Sci.\ USA
{\bf 96}, 12240 (1999).

\bibitem{kim04b}
M.J.\ Kim, T.R.\ Powers, Phys.\ Rev.\ E {\bf 69}, 061910 (2004).

\bibitem{bre63/64}
H.\ Brenner, Chem.\ Eng.\ Sci.\ {\bf 18}, 1 (1963); 
{\bf 19}, 599 (1964).

\bibitem{hin95}
K.\ Hinsen, Comput.\ Phys.\ Commun.\ {\bf 88}, 327 (1995).

\bibitem{kloe99}
P.E.\ Kloeden, E.\ Platen, {\it Numerical Solution of Stochastic
Differential Equations} (Springer, Berlin, 1999).

\bibitem{foot:kin_rev}
The same result with similar methods was obtained by Kim and
Powers in Ref.~\cite{kim04b}. They use the concept of kinematic 
reversibility of Stokes flow that manifests itself, e.g., in 
Eq.~(\ref{eq:chi'=mu*D}). That means, to arrive at 
Eq.~(\ref{eq:symm_x}), we use this concept when the torque on the
helices is reversed.



\bibitem{lam72}
H.\ Lamb, {\it Hydrodynamics} (Cambridge University Press, London, 1975).

\bibitem{fin72}
B.A.\ Finlayson, {\it The Method of Weighted Residulas and Variational
Principles} (Academic Press, New York, 1972).

\bibitem{arf95}
G.B.\ Arfken, H.J.\ Weber, {\it Mathematical Methods for Physicists}
(Academic Press, San Diego, 1995).

\bibitem{joh79}
R.E.\ Johnson, C.J.\ Brokaw, Biophys.\ J.\ {\bf 25}, 113 (1979).

\bibitem{rei04}
M.\ Reichert, H.\ Stark, J.\ Phys.: Condens.\ Matter {\bf 16}, S4085
(2004).

\end{thebibliography}
\end{document}